\def\slash#1{\setbox0=\hbox{$#1$}#1\hskip-\wd0\dimen0=5pt\advance
\dimen0 by-\ht0\advance\dimen0 by\dp0\lower0.5\dimen0\hbox
to\wd0{\hss\sl/\/\hss}}
\documentstyle[aps,epsf,prl]{revtex}
\begin{document}
\preprint{CPT-98/P.3620}
\title{A Constituent Quark-Meson Model for Heavy-Meson Processes}
\author{A. Deandrea}
\address{Centre de Physique Th\'eorique\footnote{Unit\'e Propre de
Recherche 7061}, CNRS Luminy, Case 907, F-13288 Marseille Cedex 9, France}
\author{N. Di Bartolomeo}
\address{Sissa, via Beirut 2-4, I-34014 Trieste, Italy}
\author{R. Gatto}
\address{D\'epartement de Physique Th\'eorique, Universit\'e de Gen\`eve,
24 quai E.-Ansermet, CH-1211 Gen\`eve 4, Switzerland}
\author{G. Nardulli and A. D. Polosa}
\address{Dipartimento di Fisica, Universit\`a di Bari and INFN Bari,
via Amendola 173, I-70126 Bari, Italy}

\date{February 1998}
\maketitle

\begin{abstract}
We describe the effective heavy meson lagrangian for S- and P-wave
heavy-light mesons in terms of a model based on meson-quark interactions,
where mesonic transition amplitudes are represented by diagrams with heavy
mesons attached to loops containing heavy and light constituent quarks.
The model is relativistic and incorporates the heavy quark symmetries.
The universal form factors of the heavy meson transition amplitudes
are calculated together with their slopes and compared to
existing data and limits. As  further applications of the model, strong
and radiative decays of $D^*$ and $B^*$ are considered. The agreement with
data is surprisingly good and shows that the model offers a viable alternative
to effective meson lagrangians which require a larger number of input
parameters.

\end{abstract}

\section{Introduction}
An effective theory for heavy mesons, implementing the heavy quark
symmetries, has been very successful at the phenomenological
level (for reviews see \cite{neurev} and references therein).
Predictions are easily obtained once the unknown
effective couplings are fixed from experimental data. Moreover such
an effective approach can be combined with chiral symmetry for
light mesons, thus
giving a simple framework for implementing the known approximate symmetries
of quantum chromodynamics (QCD) \cite{report}. The disadvantage of
such an approach is the number of free parameters, which grows very
rapidly if one tries to improve the calculations beyond leading order.

In order to go beyond the symmetry
approach, one should be able to derive the free couplings at the meson
lagrangian level from a more fundamental theory, for example directly
from QCD. This is clearly a   difficult task. In the long
run the definitive answer will come from systematic first principle
calculations, such  as in lattice QCD, but at present heavy meson physical
quantities such as spectra and form factors are still subject
to extrapolations, even if recent improvements are impressive
\cite{lat}. Moreover an alternative and intuitive way for
interpolating between QCD and an effective theory would be interesting
in itself, allowing us to understand better the underlying physics.
Obviously there is a price to pay for any simplification that may allow
to calculate the parameters of an effective heavy meson theory
without solving the nonperturbative QCD problem. Our point of view
here will be to consider a quark-meson lagrangian where transition
amplitudes are represented by diagrams with heavy mesons
attached to loops containing heavy and light constituent quarks.
It should be kept in mind that what we study here is only a model
and not full QCD. However one can hope to describe the essential part of
the QCD behavior in some energy range and extract useful information from it.

Since the model used in the present paper
is based on an effective constituent quark-meson lagrangian containing
both light and heavy degrees of freedom it is constrained
by the known symmetries of QCD in the limit
$m_Q\to\infty$. A similar model, for light quarks only,
 was pioneered, in the context
of chirally symmetric effective theories, several years ago
in \cite{manohar}. A related approach is the one based on the
extended Nambu Jona-Lasinio (NJL) models
 \cite{bij}, whose generalization to include both heavy and light quarks
has been studied in \cite{ebert}. In this approach  path
integral bosonization is used, which replaces the effective
4-quark interactions by Yukawa-type couplings of heavy and light
quarks with heavy meson fields. One may think of going even further
and try to bosonize directly QCD, but a number of approximations
are needed and one usually ends up with a non-local lagrangian.

In the following we are interested in the description of the heavy
meson interactions for the doublets with spin/parity $J^P=(1^+,0^+)$,
$(1^-,0^-)$, $(2^+,1^+)$, i.e. S- and P-wave heavy-light mesons.
These states are the object of experimental search
\cite{exp}. Our aim is therefore to extend the approach
of \cite{manohar} and \cite{ebert} to the description of the three
spin-parity doublets mentioned above.
We will consider a lagrangian at the meson-quark
level, fixing the free
parameters from data. This will allow us to deduce from a small number of
parameters the heavy meson couplings and form factors, with a considerable
reduction in the number of free parameters with respect to the effective
lagrangian written in terms of meson fields only.

The plan of the paper is as follows. In section 2 we give an outline
of the model. In Section 3 we compute some fundamental parameters of the
model, such as mass splittings and renormalization constants.
The model is used in Section 4 to obtain predictions for semileptonic
$B$ decays into negative parity and positive
parity charmed resonances; in particular
we compute the relevant universal form factors that describe such decays
in the $m_Q\to \infty$ limit and discuss the  Bjorken sum
rule \cite{Bjorken}. In section 5 we compute strong and radiative
$D^*$ and $B^*$ decays. Finally in Section 6 we draw our conclusions.

\section{The model}
To begin with we briefly review the  description of the heavy quark field in
the heavy quark effective theory (HQET) \cite{neurev}.
The heavy quark is defined as
follows: we indicate by $v_\mu$
the velocity of the hadron containing the heavy quark $Q$. This is almost
on shell and its momentum $p_Q$ can be written, introducing a
residual momentum $k$, as
\begin{equation}
p_Q= m_Q v + k.
\label{uno}
\end{equation}
One extracts the dominant part $m_Q v$ of the heavy quark momentum by
defining the new field $Q_v$
\begin{equation}
Q_v(x) = \exp(i m_Q v x) Q(x)=h_v(x) + H_v (x).
\end{equation}
The field $h_v$ is the large component field, satisfying the constraint
$\slash v h_v =  h_v$. If the quark $Q$ is exactly on shell, it is the only
term present in the previous equation. $H_v$, the small component field, is of
the order $1/m_Q$ and satisfies $\slash v H_v = - H_v$. The effective
lagrangian can be derived by integrating out the heavy fields in the QCD
generating functional \cite{mann}, however here we only show the effect
of the $m_Q \to \infty$ limit on the Feynman rules relevant to our
calculation \cite{flyisg}.
The heavy quark propagator  in the large $m_Q$ limit is:
\begin{equation}
{i\over{\slash p - m_Q}}
\simeq  {i\over {v\cdot k}}{{1+\slash v}\over 2}.
\end{equation}
The $(1+\slash v)/2$ projection operator can always be moved close to
a spinor $h_v$ satisfying $\slash v h_v = h_v$ since $\slash v$
commutes with the heavy quark-gluon vertex.

In our model the heavy quark propagator appears in the loops always
together with light quark propagators as the heavy mesons we consider
are made up of constituent heavy quark and light antiquark.
The light-quark momentum
$k$ may be taken equal to the integrated loop momentum and the heavy-quark
propagator in the heavy quark limit becomes:
\begin{equation}
{i\over {v\cdot k + \Delta}}
\end{equation}
where $\Delta$ is the difference between the quark mass and the common mass
of the heavy meson doublet.

\subsection{Heavy meson fields}

In order to implement the heavy quark symmetries in the spectrum of
physical states the wave function of a heavy meson has to be independent
of the heavy quark flavor and spin. It can be characterized
by the total angular momentum $s_\ell$ of the light degrees of freedom.
To each value of $s_\ell$ corresponds
a degenerate doublet of states with angular momentum $J=s_\ell \pm 1/2$.
The mesons $P$ and $P^*$ form the spin-symmetry
doublet corresponding to $s_\ell= 1/2$ (for charm for instance,
they correspond to $D$ and $D^*$).

The negative parity spin doublet $(P,P^*)$ can be represented by a
$4 \times 4$ Dirac matrix $H$, with one spinor index for the heavy quark
and the other for the light degrees of freedom.
These wave functions transform under a Lorentz transformation
$\Lambda$ as
\begin{equation}
H \to D(\Lambda)\; H\; D (\Lambda )^{-1}
\end{equation}
where $D( \Lambda)$ is the usual $4 \times 4$ representation of the Lorentz
group. Under a heavy quark spin transformation $S$, $H \to S H~~$,
where $S$ satisfies $[\slash v, S]=0$  to preserve the constraint
$\slash v H=H$.

An explicit matrix representation is:
\begin{eqnarray}
H &=& \frac{(1+\slash v)}{2}\; [P_{\mu}^*\gamma^\mu - P \gamma_5 ]\\
{\bar H} &=& \gamma_0 H^{\dagger} \gamma_0~~.
\end{eqnarray}
Here $v$ is the heavy meson velocity, $v^\mu P^*_{a\mu}=0$ and
$M_H=M_P=M_{P^*}$. Moreover $\slash v H=-H \slash v =H$, ${\bar H}
\slash v=-\slash v {\bar H}={\bar H}$ and $P^{*\mu}$ and $P$ are annihilation
operators normalized as follows:
\begin{eqnarray}
\langle 0|P| Q{\bar q} (0^-)\rangle & =&\sqrt{M_H}\\
\langle 0|{P^*}^\mu| Q{\bar q} (1^-)\rangle & = & \epsilon^{\mu}\sqrt{M_H}~~.
\end{eqnarray}
The formalism for higher spin states is given in \cite{falk}.
We will consider only the extension to $P$-waves of the system $Q\bar q$.
The heavy quark effective theory predicts two distinct multiplets, one
containing a $0^+$ and a $1^+$ degenerate state, and the other one a $1^+$ and
a $2^+$ state. In matrix notations, analogous to the ones used for the
negative
parity states, they are described by
\begin{equation}
S={{1+\slash v}\over 2}[P_{1\mu}^{*\prime} \gamma^\mu\gamma_5-P_0]
\end{equation}
and
\begin{equation}
T^\mu={\frac {1+\slash v}{2}}\left[P_2^{* \mu\nu}\gamma_\nu-\sqrt{\frac 3
2}
P^*_{1\nu}\gamma_5\left(g^{\mu\nu}-\frac 1 3
\gamma^\nu(\gamma^\mu-v^\mu)\right)
\right].
\end{equation}
These two multiplets have $s_\ell=1/2$ and $s_\ell=3/2$ respectively, where
$s_\ell$ is conserved together with the spin $s_Q$ in the infinite quark
mass limit because $\vec J={\vec s}_\ell+{\vec s}_Q$.

\subsection{Interaction terms}

The effective quark-meson lagrangian we introduce contains two terms:
\begin{equation}
{\cal L} ~=~{\cal L}_{\ell \ell}~+~{\cal L}_{h \ell}~.\label{lagra}
\end{equation}
The first term involves only the light degrees of freedom,
i.e. the light quark fields $\chi$ and the pseudo-scalar
$SU(3)$ octet of mesons $\pi$. At the lowest order one has:
\begin{eqnarray}
{\cal L}_{\ell \ell}&=&{\bar \chi} (i  D^\mu \gamma_\mu +g_A
{\cal A}^\mu \gamma_\mu \gamma_5) \chi - m {\bar \chi}\chi \nonumber\\
&+& {f_{\pi}^2\over 8} \partial_{\mu} \Sigma^{\dagger} \partial^{\mu}
\Sigma .
\end{eqnarray}
Here $D_\mu = \partial_\mu-i {\cal V}_\mu$,
$\xi=\exp(i\pi/f_\pi )$, $\Sigma=\xi^2$, $f_\pi=130$ MeV and
\begin{eqnarray}
{\cal V}^\mu &=& {1\over 2} (\xi^\dagger \partial^\mu \xi +\xi \partial^\mu
\xi^\dagger) \nonumber \\
{\cal A}^\mu &=& {i\over 2} (\xi^\dagger \partial^\mu \xi -\xi \partial^\mu
\xi^\dagger)~. \label{av}
\end{eqnarray}
The term with $g_A$ is the coupling of pions to light quarks; it will not
be used in the sequel. It is a free parameter, but in NJL model $g_A=1$.
Differently from \cite{manohar} the gluon field is absent in the present
description. Apart from the mass term ${\cal L}_{\ell \ell}$ is chiral 
invariant.

We can introduce a quark-meson effective lagrangian
involving heavy and light quarks and heavy mesons. At lowest order we have:
\begin{eqnarray}
{\cal L}_{h \ell}&=&{\bar Q}_v i v\cdot \partial Q_v
-\left( {\bar \chi}({\bar H}+{\bar S}+ i{\bar T}_\mu
{D^\mu \over {\Lambda_\chi}})Q_v +h.c.\right)\nonumber \\
&+&\frac{1}{2 G_3} {\mathrm {Tr}}[{\bar H}H]+\frac{1}{2 G^{\prime}_3}
{\mathrm {Tr}}[{\bar S}S]
+\frac{1}{2 G_4}
{\mathrm {Tr}} [{\bar T}_\mu T^\mu]  \label{qh}
\end{eqnarray}
where the meson fields $H,~S,~T$ have been defined
before, $Q_v$ is the effective
heavy quark field, $G_3,~G^{\prime}_3,~G_4$ are coupling constants (to
allow a comparison with previous work we follow, as far as
possible, the notations of \cite{ebert})
and $\Lambda_\chi$ ($= 1$ GeV) has been introduced for dimensional reasons. 
Lagrangian (\ref{qh}) has heavy spin and flavor symmetry. 
This lagrangian comprises three
terms containing respectively $H$, $S$ and $T$.
~From a theoretical point of view, the $G_3$ and $G^{\prime}_3$ couplings
could have been determined by $\Delta_H$ and $\Delta_S$, the experimental 
values of the mass differences
between the mesons, belonging to the H and S multiplets respectively, and their
heavy quark constituent masses. Since these data are not all known,
we have adopted the dynamical information coming from the
NJL-model as proposed in~\cite{ebert} and
so we use the following modified version of (\ref{qh}) in which
we give the fields $H$ and $S$ the same coupling to
the quarks:

\begin{eqnarray}
{\cal L}_{h \ell}&=&{\bar Q}_v i v\cdot \partial Q_v
-\left( {\bar \chi}({\bar H}+{\bar S}+ i{\bar T}_\mu
{D^\mu \over {\Lambda_\chi}})Q_v +h.c.\right)\nonumber \\
&+&\frac{1}{2 G_3} {\mathrm {Tr}}[({\bar H}+{\bar S})(H-S)]
+\frac{1}{2 G_4}
{\mathrm {Tr}} [{\bar T}_\mu T^\mu ] \label{qh1}
\end{eqnarray}

In doing this we assume that this effective quark meson
lagrangian can be justified as a remnant of a four quark interaction of
the NJL type by partial bosonization. $H$ and $S$ are degenerate in mass
in the light-sector chirally symmetric phase. In the broken phase the mass
splitting
is calculable and will be computed in the next section in terms of the
order parameter $m$, the constituent light quark mass. The part containing the
$T$ field can not be extracted from a bosonized NJL contact interaction and
requires a new coupling constant $G_4$.

An essential ingredient of the nonperturbative behavior of QCD is the
suppression of large momentum flows through light-quark lines in the
loops. The way this suppression is introduced in the model is crucial
and is part of the definition of the model itself. One can for example
include factors in the vertices which damp the loop integrals when the
light-quark momentum is larger than some scale of the order of
1 GeV\cite{holdom}.
Our model describes the interactions in terms of effective vertices between a
light quark, a heavy quark and a heavy meson; we describe the heavy
quarks and heavy mesons consistently with HQET,
and thus the heavy quark propagator in the loop contains the
residual momentum $k$ which arises from the interaction with the light
degrees of freedom. It is therefore natural to assume an ultraviolet cut-off
on the loop momentum of the order of $\Lambda \simeq 1$ GeV, even
if the  heavy quark mass is larger than the cut-off.

Concerning the infrared behavior, the model is not confining and thus
its range of validity can not be extended below energies of the order of
$\Lambda_{QCD}$. In practice one introduces an infrared cut-off $\mu$, in order
to drop the unknown confinement part of the quark interaction.

We choose a proper time regularization (a different choice is followed
in \cite{bardeen}). After continuation to
the Euclidean space it reads, for the light quark propagator:
\begin{equation}
\int d^4 k_E \frac{1}{k_E^2+m^2} \to \int d^4 k_E \int^{1/
\Lambda^2}_{1/\mu^2} ds\; e^{-s (k_E^2+m^2)}\label{cutoff}
\end{equation}
where $\mu$ and $\Lambda$ are
 infrared and  ultraviolet   cut-offs.
This choice of regularization is consistent with the one in the second
paper in \cite{ebert}. In (\ref{cutoff}) $m$ is the constituent light quark
mass, whose non vanishing value implies the mass splitting between the $H$-type
and $S$-type multiplets. Its value, as well as the values of the cutoffs
$\mu$ and $\Lambda$ are in principle adjustable parameters.
On physical grounds, we expect that $\Lambda\simeq 1$ GeV, $m\simeq \mu \simeq
10^2$ MeV. Given our choice of the cut-off prescription as in \cite{ebert},
we assume for $\Lambda$ the same value
used there i.e. $\Lambda=1.25$ GeV; we observe, however, that
the numerical outcome of the subsequent
calculations are not strongly dependent on the value of $\Lambda$. The
constituent mass $m$ in the NJL models
represents the order
parameter discriminating between the phases of broken and unbroken chiral
symmetry and can be fixed by solving a gap equation, which gives $m$ as a
function of the scale mass $\mu$ for given values of the other parameters. In
the second paper of
ref. \cite{ebert} the values $m=300$ MeV and $\mu=300$ MeV are used
and we shall assume the same values. As shown there, for smaller
values of $\mu$, $m$ is constant (=300 MeV) while for
much larger values of $\mu$, it  decreases and in
particular it vanishes for $\mu=550$ MeV.

\section{Renormalization constants and masses}

For the computation of the constants $G_3$, $G_4$ appearing in the 
quark-meson effective lagrangian of Eq.(\ref{qh})
we consider the meson self energy diagram depicted in Fig.1. 

\begin{figure}
\epsfxsize=8truecm
\centerline{\epsffile{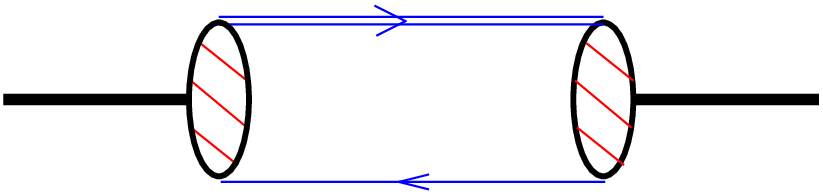}}
\noindent
{\bf Fig. 1} - {Self-energy diagram for the heavy meson field.}
\end{figure}

For $H$ and $S$ states the evaluation of Fig.1 gives the results ($k$ is
the residual momentum, see eq. (\ref{uno})):
\begin{equation}
{\mathrm {Tr}} [-{\bar H}\; \Pi_H(v\cdot k) \; H ]
=i N_c\int\frac{d^4 \ell}{(2 \pi)^4} \frac{{\mathrm {Tr}}[(\slash \ell -
\slash k -m){\bar H}H]}{ ((\ell-k)^2 -
m^2+i\epsilon)(v\ell +i \epsilon)} \label{pih}
\end{equation}
and
\begin{equation}
{\mathrm {Tr}} [+{\bar S}\; \Pi_S(v\cdot k) \; S ]
=i N_c\int\frac{d^4 \ell}{(2 \pi)^4}
\frac{{\mathrm {Tr}}[(\slash \ell - \slash k +m){\bar S}S]} {((\ell-k)^2
-m^2+i\epsilon)(v\ell +i \epsilon)}\label{pis}
\end{equation}
respectively, and for $T$ states we get
\begin{equation}
{\mathrm {Tr}}[+{\bar T}_\mu \; \Pi_T(v\cdot k) \; T^\mu ] 
=\frac{i N_c}{\Lambda^2_\chi}\int\frac{d^4 \ell}{(2 \pi)^4}
\frac{{\mathrm {Tr}}[(\ell_\mu -k_\mu) (\slash \ell -\slash k +m)
(\ell_\nu- k_\nu){\bar T}^\mu T^\nu ] }{((\ell-k)^2
-m^2+i\epsilon)(v\ell +i \epsilon)}\label{pit}
\end{equation}

We notice that the only difference between $\Pi_H$ and $\Pi_S$ is in the
sign in front of $m$ in the numerators of the integrands.
We now expand (\ref{pih}),(\ref{pis}) and(\ref{pit}) around  $\Delta_H$,
$\Delta_S$ and $\Delta_T$ respectively:
\begin{equation}
\Pi(v\cdot k)\simeq \Pi(\Delta)+ \Pi^\prime(\Delta)(v\cdot k - \Delta)~.
\end{equation}
In this way we obtain the
kinetic part of the effective lagrangian for heavy mesons in
the
usual form \cite{report}
\begin{eqnarray}
{\cal L}_{eff}&=& {\mathrm {Tr}} {\bar {\hat H}}(i v\cdot \partial
-\Delta_H){\hat H}
+{\mathrm {Tr}} {\bar {\hat S}}(i v\cdot\partial -\Delta_S) {\hat S}
\nonumber \\
&+&{\mathrm {Tr}}{\bar {\hat T}}_\mu(i v\cdot\partial -\Delta_T)
{\hat T}^\mu~,
\end{eqnarray}
provided we satisfy the  conditions:
\begin{eqnarray}
\frac{1}{2 G_3}&=&\Pi_H(\Delta_H)=\Pi_S(\Delta_S) \label{g3}\\
\frac{1}{2 G_4}&=&\Pi_T(\Delta_T)
\label{g4}
\end{eqnarray}
and renormalize the fields as :
\begin{eqnarray}
{\hat H} &=&  \frac{H}{\sqrt {Z_H}} \\
{\hat S} &=&  \frac{S}{\sqrt {Z_S}} \\
{\hat T} &=& \frac{T}{\sqrt {Z_T}}~.
\end{eqnarray}
 The renormalization constants are given by:
\begin{equation}
Z_j^{-1}=\Pi^{\prime}(\Delta_j)~.
\end{equation}
Here, prime denotes differentiation and
 $j=H,~S$ and $T$. In the previous equations,
 $\Delta_H,~\Delta_S,~\Delta_T$
are the mass differences between the heavy mesons $H,S,T$ and the heavy quark.
The expressions of $\Pi_H$, $\Pi_S$, and $\Pi_T$ are:
\begin{eqnarray}
\Pi_H (\Delta_H) &=& I_1 + ( \Delta_H + m) I_3(\Delta_H)\\
\Pi_S (\Delta_S) &=& I_1 + (\Delta_S - m)I_3(\Delta_S)\\
\Pi_T (\Delta_T) &=& {1\over {\Lambda_{\chi}^2}}
\left[ -{I'_1\over 4}+
\frac{m+\Delta_T}{3} [I_0(\Delta_T) + \Delta_T I_1 +
(\Delta_T^2-m^2) I_3(\Delta_T)] \right]
\end{eqnarray}
while the field renormalization constants are :
\begin{equation}
 Z_H^{-1} = (\Delta_H+m) {\frac{\partial I_3(\Delta_H)}{\partial \Delta_H}}
 +I_3(\Delta_H)\label{zh}
\end{equation}

\begin{equation}
 Z_S^{-1} = (\Delta_S-m) {\frac{\partial I_3(\Delta_S)}{\partial \Delta_S}}
 +I_3(\Delta_S)
\end{equation}

\begin{eqnarray}
 Z_T^{-1} &=&\frac{1}{3 \Lambda^2_\chi} \Big[ (\Delta_T^2-m^2) [(m+\Delta_T)
{\frac{\partial I_3(\Delta_T)}{\partial \Delta_T}} +I_3(\Delta_T)]\nonumber \\
&+& (m+\Delta_T)
[{\frac{\partial I_0(\Delta_T)}{\partial \Delta_T}}+ I_1
+2 \Delta_T I_3(\Delta_T)] +I_0 +\Delta_T I_1 \Big]
\end{eqnarray}
The integrals $I_0,I_1,I'_1$ and $I_3$ are defined in the Appendix.
We employ  the first equation in (\ref{g3}) to obtain $G_3$ from a given value
of $\Delta_H$ (we use $\Delta_H$ in the range $0.3-0.5$ GeV), while the
second equation in (\ref{g3}) can be used to determine $\Delta_S$.

For $\Delta_T$ we can use experimental information.
Let us call  $M_H$ and $M_T$ the spin-averaged $H$-multiplet
and $T$-multiplet masses. We can write
$M_H=m_Q+\Delta_H+\Delta^\prime_H/m_Q$ and
$M_T=m_Q+\Delta_T+\Delta^\prime_T/m_Q$. For charm, the
positive parity experimental masses and widths
are as follows: for $D_2^*$ we have $m_{D^{*}_2(2460)^0}=2458.9\pm 2.0$ MeV,
$\Gamma_{D^{*}_2(2460)^0}=23\pm 5$ MeV and
$m_{D^{*}_2(2460)^{\pm}}=2459\pm 4$ MeV,
$\Gamma_{D^{*}_2(2460)^\pm}=25^{+8}_{-7}$  MeV. These particle are
identified  with the state $2^+$ belonging to the $T$
$\frac{3}{2}^+$ multiplet.
As for the $D^{*}_1(2420)$ state, experimentally we have
$m_{D^{*}_1(2420)^0}=2422.2\pm 1.8$ MeV,
$\Gamma_{D^{*}_1(2420)^0}=18.9^{+4.6}_{-3.5}$ MeV; this
particle can be identified  with the state $1^+$
belonging to the $T$ multiplet, neglecting a possible small mixing between the
two $1^+$ states belonging to the $S$ and $T$ multiplets \cite{report}.
We observe that the narrowness of the states
$D_2^*$ and $D_1^*$ is due to the fact that their strong decays
occur via $D$-waves. From this analysis we get
$\Delta_T - \Delta_H +(\Delta^\prime_T - \Delta^\prime_H)/m_c\simeq 470$ MeV.
For beauty, the experimental data on positive parity resonances
show a bunch of states with a mass $M_{B^{**}}=5698\pm 12$ MeV and
width $\Gamma=128\pm 18$ MeV \cite{PDG}. If we identify this mass with the
mass of the narrow $T$ states, we obtain
$\Delta_T - \Delta_H +(\Delta^\prime_T - \Delta^\prime_H)/m_b\simeq 380$ MeV.
For reasonable values of the heavy quark masses, we get
\begin{equation}
\Delta_T - \Delta_H \simeq 335 \;{\mathrm {MeV}}\;,
\end{equation}
which is the value we adopt.
The preceding analysis produces the
figures in Table I.
\begin{table} [htb]
\hfil
\vbox{\offinterlineskip
\halign{&#& \strut\quad#\hfil\quad\cr
\tableline
\tableline
&$\Delta_H$ && $\Delta_S$ && $\Delta_T$& \cr
\tableline
&$0.3$&& $0.545$&& $0.635$ &\cr
&$0.4$&& $0.590$&& $0.735$ &\cr
&$0.5$&& $0.641$&& $0.835$ &\cr
\tableline
\tableline}}
\caption{$\Delta$ values (in GeV)}
\end{table}
In Table II we report the computed values of $G_j$ and $Z_j$
for three values of $\Delta_H$.

Let us finally observe that we predict, according to the value of $\Delta_H$,
a value for the $S-$multiplet mass $m=2165 \pm 50$ MeV;
these states, called in the literature $D_0,D^{*\prime}_1$ have not been
observed yet since they are expected to be rather broad (for instance in
\cite{QCD} one predicts $\Gamma(D_0\to D^+\pi^-)\simeq 180$ MeV
and  $\Gamma(D_1^{*\prime}\to D^{*+}\pi^-)\simeq 165$ MeV). Theoretical
predictions in the literature are somehow larger ($m \simeq 2350$ MeV).
\begin{table} [htb]
\hfil
\vbox{\offinterlineskip
\halign{&#& \strut\quad#\hfil\quad\cr
\tableline
\tableline
&$\Delta_H $&& $1/G_3 $&& $Z_H $&& $Z_S $&& $Z_T $&& $1/G_4$ &\cr
\tableline
&$0.3$&& $0.16$&&  $4.17$&& $1.84$&& $2.95$&&  $0.15$ & \cr
&$0.4$&& $0.22$&&  $2.36$&& $1.14$&& $1.07$&&  $0.26$ & \cr
&$0.5$&& $0.345$&& $1.142$&& $0.63$&& $0.27$&& $0.66$ & \cr
\tableline
\tableline}}
\caption{Renormalization constants and couplings.
$\Delta_H$ in GeV; $G_3$, $G_4$ in GeV$^{-2}$,
$Z_j$ in GeV$^{-1}$.}
\end{table}

\section{Semileptonic decays and form factors}

To begin with, we compute the leptonic decay constants ${\hat F}$
and ${\hat F}^+$
that are defined as follows:
\begin{eqnarray}
\langle 0|{\bar q}
\gamma^\mu \gamma_5 Q  |H( 0^-, v)\rangle  & =&i \sqrt{M_H} v^\mu
{\hat F}\\
\langle 0|{\bar q}
\gamma^\mu  Q |S( 0^+, v)\rangle & =&i \sqrt{M_S} v^\mu {\hat F}^+~~.
\end{eqnarray}
It is easy to compute these constants by a loop calculation
similar to that considered in the
previous section for the  self-energy ; one finds

\begin{eqnarray}
{\hat F}&=&\frac{\sqrt{Z_H}}{G_3}\\
{\hat F}^+ &=& \frac{\sqrt{Z_S}}{G_3}
\end{eqnarray}

The numerical results for different values of the
parameters can be found in Table III.
\begin{table} [htb]
\hfil
\vbox{\offinterlineskip
\halign{&#& \strut\quad#\hfil\quad\cr
\tableline
\tableline
&$\Delta_H$&& ${\hat F}$&& ${\hat F}^+$&\cr
\tableline
&$0.3$&& $0.33$&& $0.22$&\cr
&$0.4$&& $0.34$&& $0.24$&\cr
&$0.5$&& $0.37$&& $0.27$&\cr
\tableline
\tableline}}
\caption{${\hat F}$ and ${\hat F}^+$ for
various values of $\Delta_H $. $\Delta_H$ in GeV, leptonic constants
in GeV$^{3/2}$.}
\end{table}
We observe that, neglecting logarithmic corrections,
${\hat F}$ and
${\hat F}^+$ are related,
in the infinite heavy quark mass limit, to the
leptonic decay constant $f_B$ and $f^+$
defined by
\begin{eqnarray}
\langle 0|{\bar q} \gamma^\mu \gamma_5 b  |B(p)\rangle
&=& i  p^\mu f_B \\
\langle 0|{\bar q} \gamma^\mu  b  |B_0(p)\rangle
&=& i  p^\mu f^+
\end{eqnarray}
by the relations $f_B={\hat F}/\sqrt{M_B}$
and $f^+={\hat F}^+/\sqrt{M_{B_0}}$.
For example, for $\Delta_H=400$ MeV, we obtain from Table III:
\begin{eqnarray}
f_B &\simeq& 150 \;{\mathrm {MeV}}\\
f^+ &\simeq& 100 \;{\mathrm {MeV}}\;.
\end{eqnarray}

The numerical values that can
be found in the literature agree with our results. For example the QCD sum
rule analysis of \cite{neubert} gives ${\hat F}=0.30 \pm 0.05$ GeV$^{3/2}$
(without $\alpha_s$ corrections) and higher values
(around $0.4-0.5$ GeV$^{3/2}$) including radiative corrections.
As for lattice QCD, in \cite{lat} the
value summarizing the present status of lattice calculations for
$f_B$ is $170\pm 35$ MeV.
As for $F^+$, a QCD sum rule analysis \cite{CNP}
gives $F^+=0.46\pm 0.06$ GeV$^{3/2}$, which is significantly higher
than the results reported in Table III.

Let us now consider semileptonic decays. The first
analysis to be performed is the study  of the Isgur-Wise function $\xi$
which is defined by:
\begin{eqnarray}
\langle D(v^\prime)|{\bar c} \gamma_\mu (1-\gamma_5) b|
B(v)\rangle  & =& \sqrt{M_B M_D} C_{cb} \;
\xi(\omega) (v_\mu + v^{\prime }_{\mu}) \\
\langle D^*(v^\prime, \epsilon)|{\bar c} \gamma_\mu (1- \gamma_5) b| B(v)
\rangle  & =&
\sqrt{M_B M_{D^*}} C_{cb}\; \xi(\omega)[i \epsilon_{\mu \nu \alpha \beta}
\epsilon^{*\nu}v^{\prime \alpha}v^\beta  \nonumber \\
&-& (1+\omega)\epsilon^*_\mu+ (\epsilon^*\cdot
v)v^\prime_\mu]
\end{eqnarray}
where $\omega= v \cdot v^\prime$ and $C_{cb}$
is a coefficient containing
logarithmic corrections depending on $\alpha_s$; within our approximation
it can be put equal to 1: $C_{cb}=1$. We also note that,
in the leading order we are considering here $\xi(1)=1$.
To compute $\xi$ one has to evaluate the diagram of Fig.2.

\begin{figure}
\epsfxsize=8truecm
\centerline{\epsffile{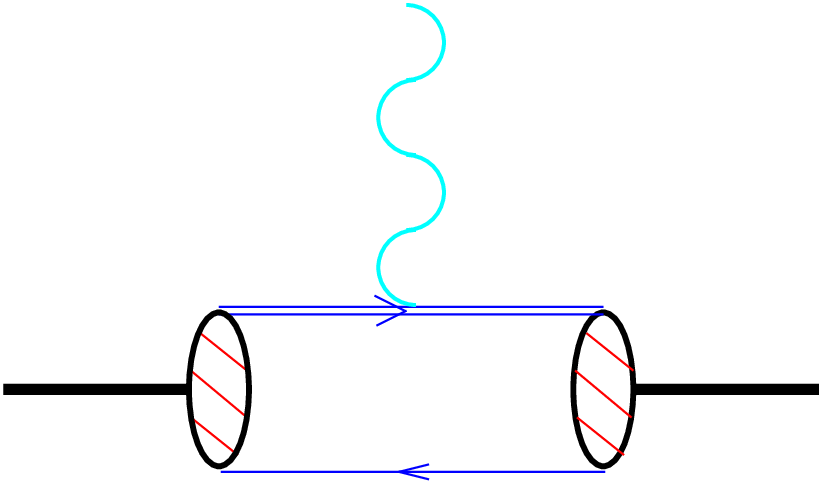}}
\noindent
{\bf Fig. 2} - {Weak current insertion on the heavy quark line for the
heavy meson form factor calculation.}
\end{figure}

One finds \cite{ebert}:
\begin{equation}
\xi(\omega)=Z_H \left[ \frac{2}{1+\omega} I_3(\Delta_H)+\left( m+\frac{2
\Delta_H}{1+\omega} \right)
I_5(\Delta_H, \Delta_H,\omega) \right] ~~.\label{xi}
\end{equation}
\begin{figure}
\epsfxsize=8truecm
\centerline{\epsffile{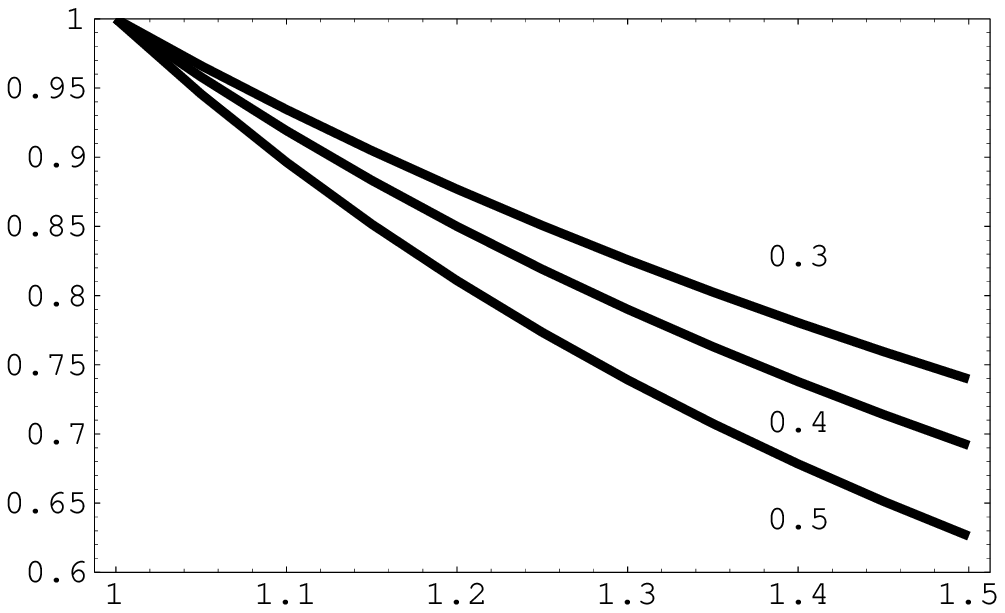}}
\noindent
{\bf Fig. 3} - {Isgur-Wise form factor at different $\Delta$ values.}
\end{figure}
The $\xi$ function is plotted in Fig.3.

The integrals $I_3,~I_5$ can be found in the Appendix.
It is obvious from (\ref{zh}),(\ref{xi}) that $\xi(\omega)$
is correctly normalized, i.e. $\xi(1)=1$. For
the slope of the Isgur-Wise function, defined by
\begin{equation}
\rho^2_{IW}=-\frac{d\xi}{d\omega}(1)~~,
\end{equation}
one gets the values reported in Table IV.
The plot of the Isgur-Wise function is given in fig. 3 for three
values of $\Delta_H$. It is useful to compare our results with other
approaches.
QCD sum rules calculations are in the range $\rho^2_{IW}= 0.54-1.0$, therefore
they agree with our results\cite{Blok}. The results from different quark models
\cite{godfrey},\cite{veseli},\cite{CCCN},\cite{ISGW}
have been recently reviewed by the authors of \cite{oliver}; they
have studied the results of different  models with the
common approach of boosting the wave-functions by the Bakamjian-Thomas method
\cite{BT}; for $ \rho^2_{IW}$ they obtain results in the range $0.97-1.28$,
to be compared to the results in table 4. Lattice QCD gives
significantly smaller results; for example in \cite{Chris} the value
$\rho^2_{IW}(1)=0.64$ is obtained.

Let us now turn to the  form factors describing the semileptonic
decays of a meson belonging to the fundamental negative
parity multiplet $H$ into the
positive parity mesons in the $S$ and $T$ multiplets.
Examples of these decays are
\begin{equation}
B \rightarrow D^{**} \ell \nu \label{bd**}
\end{equation}
where $D^{**}$ can be either a $S$ state
(i.e. a $0^+$ or $ 1^+$ charmed meson having
$s_{\ell}=1/2$)
or  a $T$ state ( i.e. a $2^+$ or $ 1^+$ charmed meson having $s_{\ell}=3/2$).

The decays in (\ref{bd**}) are described  by two form
factors $\tau_{1/2}, \tau_{3/2}$
\cite{IW2} according to
\begin{eqnarray}
&&<D_2^*(v^{\prime},\epsilon)| {\bar c} \gamma_\mu (1- \gamma_5) b| B(v)>
=\sqrt{ 3 M_B M_{D_2^*}} ~\tau_{3/2}(\omega)\times
\nonumber\\
&&\left[ i  \epsilon_{\mu  \alpha \beta \gamma }
\epsilon^{*\alpha \eta} v_\eta v^{\prime
\beta}v^\gamma -
 [( \omega+1) \epsilon^*_{\mu \alpha} v^\alpha -
\epsilon^*_{\alpha \beta} v^\alpha v^\beta v^\prime_\mu]\right]
\label{tau32}
\end{eqnarray}
\begin{eqnarray}
&&<D_1^*(v^{\prime},\epsilon)| {\bar c}
\gamma_\mu (1- \gamma_5) b| B(v)>
=\sqrt{\frac{M_B M_{D_1^*}}{2}} ~\tau_{3/2}(\omega)\times
\nonumber\\
&&\left\{ (\omega^2-1) \epsilon_{\mu}^*
+  ( \epsilon^*\cdot v) [3~v_\mu -(\omega-2)
v^\prime_\mu ]-  i (\omega+1) \epsilon_{\mu
\alpha \beta \gamma} \epsilon^{* \alpha}
v^{\prime \beta} v^\gamma \right\}
\label{tau32bis}
\end{eqnarray}

\begin{eqnarray}
\langle D_0(v^\prime)|{\bar c} \gamma_\mu (1-\gamma_5) b| B(v)\rangle
& =& \sqrt{M_B M_{D_0}}~2~
\tau_{1/2}(\omega) (  v^{\prime }_{\mu}- v_\mu) \\
\langle D_1^{*\prime}(v^\prime, \epsilon)|{\bar c} \gamma_\mu (1- \gamma_5) b|
 B(v)\rangle
& =&
\sqrt{M_B M_{D_1^{*\prime}}}~\tau_{1/2}(\omega)\{
2~i~\epsilon_{\mu  \alpha \beta \gamma}
\epsilon^{*\alpha}v^{\prime \beta}v^\gamma \nonumber\\
+2[(1-\omega)\epsilon^*_\mu+ (\epsilon^*\cdot v)v^\prime_\mu]\}~.
\end{eqnarray}
In all these equations we neglect logarithmic corrections.
The form factors $\tau_{1/2}(\omega),~\tau_{3/2}(\omega)$
can be computed by a loop calculation similar to the one used to obtain
$\xi(\omega)$. The result is
\begin{equation}
\tau_{1/2}(\omega)=\frac{\sqrt{Z_H Z_S}}{2(1-\omega)}
\left[ I_3(\Delta_S)-
I_3(\Delta_H)
+\left(\Delta_H - \Delta_S + m(1-\omega) \right)
I_5(\Delta_H, \Delta_S,\omega) \right]
\end{equation}
and

\begin{eqnarray}
\tau_{3/2}(\omega)&=&-{{\sqrt{Z_H\,Z_T}}\over{{\sqrt{3}}}} \,
     \Big[m \Big( {{I_3(\Delta_H) - I_3(\Delta_T) -
              \left( \Delta_H - \Delta_T \right) \,
               I_5(\Delta_H,\Delta_T,\omega )}\over
            {2\,\left( 1 - \omega  \right) }} \nonumber \\
&-&{{I_3(\Delta_H) + I_3(\Delta_T)  +
              \left( \Delta_H + \Delta_T \right)  \,
               I_5(\Delta_H,\Delta_T,\omega )}\over
            {2\,\left( 1 + \omega  \right) }} \Big) \nonumber\\
&-& {{1}\over {2\,\left( -1 - \omega  + {{\omega }^2} +  {{\omega }^3}
\right) }} \Big( -3\,S(\Delta_H,\Delta_T, \omega ) -
           \left( 1 - 2\,\omega   \right) \,S(\Delta_T,\Delta_H,
\omega )\nonumber  \\
&+& (1-{{\omega }^2}) \,T(\Delta_H,\Delta_T,\omega  ) -
   2\, (1 - 2\,\omega ) \,U(\Delta_H,\Delta_T,\omega ) \Big)  \Big]
\end{eqnarray}
where the integrals $S,T,U$ are defined in the appendix.

Given the small phase space which is available for these decays
($\omega_{max}=1.33$ for $D_1^*,~D_2^*$
and $\omega_{max}\simeq 1.215$ for $D_1^{*\prime},~D_0$),
we can approximate
\begin{equation}
\tau_j(\omega)\simeq\tau_j(1)\times [ 1-\rho^2_j(\omega -1)]~.
\end{equation}

Numerically we find the results reported in Table IV.
\begin{table} [htb]
\hfil
\vbox{\offinterlineskip
\halign{&#& \strut\quad#\hfil\quad\cr
\tableline
\tableline
&$\Delta_H$&& $\xi(1)$ && $\rho^2_{IW}$&&
$\tau_{1/2}(1)$&& $\rho^2_{1/2}$&& $\tau_{3/2}(1)$&&
$\rho^2_{3/2}$&\cr
\tableline
&$0.3$&&  $1$&& $0.72$&& $0.08$&& $0.8$&& $0.48$&& $1.4$&\cr
&$0.4$&&  $1$&& $0.87$&& $0.09$&& $1.1$&& $0.56$&& $2.3$&\cr
&$0.5$&&  $1$&& $1.14$&& $0.09$&& $2.7$&& $0.67$&& $3.0$&\cr
\tableline
\tableline}}
\caption{Form factors and slopes. $\Delta_H$ in GeV.}
\end{table}
An important test of our approach is represented by the
Bjorken sum rule, which states
\begin{equation}
\rho^2_{IW}=\frac{1}{4}+\sum_k \left[|\tau_{1/2}^{(k)}(1)|^2~+~
2|\tau_{3/2}^{(k)}(1)|^2\right]~.
\end{equation}
Numerically we find that the first excited resonances, i.e. the
$S$ and $T$ states ($k=0$) practically saturate
the sum rule for all the three values of $\Delta_H$.

In Table V we compare our results (for $\Delta_H=0.4 $
GeV) with other approaches. For $\tau_{3/2}$ we find a broad 
agreement with some of the constituent quark model results, whereas for 
$\tau_{1/2}$ we only agree with \cite{defazio2}.
\begin{table} [htb]
\hfil
\vbox{\offinterlineskip
\halign{&#& \strut\quad#\hfil\quad\cr
\tableline
\tableline
&$\tau_{1/2}(1)$&& $\rho^2_{1/2}$&& $\tau_{3/2}(1)$&&
$\rho^2_{3/2}$&& Ref. &\cr
\tableline
&$0.09$&& $1.1$&& $0.56$&& $2.3$&& This work&\cr
&$0.41$&& $1.0$&& $0.41(input)$&& $1.5$&& \cite{LLSW} &\cr
&$0.25$&& $0.4$&& $0.28$&& $0.9$&& \cite{CNP}&\cr
&$0.31$&& $2.8$&& $0.31$&& $2.8$&& \cite{ISGW}&\cr
&$0.41$&& $1.4$&& $0.66$&& $1.9$&& \cite{wambach}&\cr
&$0.059$&& $0.73$&& $0.515$&& $1.45$&& \cite{oliver},\cite{CCCN}&\cr
&$0.225$&& $0.83$&& $0.54$&& $1.50$&& \cite{oliver},\cite{godfrey}&\cr
\tableline
\tableline}}
\caption{Parameters of the form factors $\tau_{1/2}$, $\tau_{3/2}$. The
results in this table are for $\Delta_H = 0.4$ GeV.}
\end{table}
Finally in Table VI we present our results for the branching ratios of
B semileptonic decays to $S-$ and $P-$wave charmed mesons for three values of
$\Delta_H$ computed with $V_{cb}=0.038$ \cite{neub} and $\tau_{B}=1.62$
psec. We see that data favor a value of $\Delta_H\simeq 400-500$ MeV.
\begin{table} [htb]
\hfil
\vbox{\offinterlineskip
\halign{&#& \strut\quad#\hfil\quad\cr
\tableline
\tableline
&Decay mode&& $\Delta_H=0.3$&& $\Delta_H=0.4$&& $\Delta_H=0.5$&&
Exp. & \cr
\tableline
&${B}\to D\ell\nu $&& $3.0$&&$2.7$&&$2.2$&&$1.9 \pm 0.5$
\cite{PDG}& \cr
&${B}\to D^{*}\ell\nu$&& $7.6$&& $6.9$&& $5.9$&& $4.68\pm 0.25$
\cite{PDG} &\cr
&${B}\to D_0\ell\nu$&& $0.03$&& $0.005$&& $0.003$&& -- &\cr
&${B}\to D^{*\prime}_{1}\ell\nu$&& $0.03$&& $0.008$&& $0.0045$&&-- &\cr
&${B}\to D_{1}^*\ell\nu$&& $0.27$&& $0.18$&& $0.13$&& $0.74 \pm 0.16$
\cite{aleph}&\cr
&${B}\to D^{*}_{2}\ell\nu$&& $0.43$&& $0.34$&& $0.30$&& $<0.85$&\cr
\tableline
\tableline}}
\caption{Branching ratios (\%) for semileptonic  $B$ decays.
Theoretical predictions for three
values of $\Delta_H$ and experimental results (for $B^0$ decays).
Units of $\Delta_H$ in GeV.}
\end{table}

\section{Strong and Radiative decays of heavy mesons}

In this section we consider the strong decays
\begin{eqnarray}
&H& \rightarrow H \pi \label{HHpi}\\
&S& \rightarrow H\pi \label{SHpi}
\end{eqnarray}
as well as the radiative decay
\begin{eqnarray}
P^* \rightarrow P \gamma
\end{eqnarray}
where $P^*$ and $P$ are the $1^-$ and $0^-$ members of the multiplet $H$.

\subsection{Strong decays}
The calculation of the strong coupling constants describing the decays
$H \rightarrow H\pi$ (i.e. $ D^*
\rightarrow D\pi$) and $S \rightarrow H
\pi$  is strongly simplified  by
adopting the soft pion limit. In the case of the decay
$D^* \rightarrow D\pi$  this procedure introduces only a small error since
the phase space is actually very small (for the
$S \rightarrow H\pi$ decay the situation is different, see below).

Let us define
$g_{D^{*}D\pi}$ by the equation:
\begin{equation}
<\pi^+ (q) D^0(p)|D^{*+}(p^{\prime},\epsilon)>=ig_{D^* D \pi}
\epsilon^{\mu}q_{\mu}~.
\end{equation}
The constant
$g_{D^{*}D\pi}$ is related to the strong coupling constant
of the effective meson field theory $g$ appearing in \cite{report}
\begin{equation}
{\cal L}=ig {\mathrm {Tr}}(\overline{H} H \gamma^{\mu}\gamma_5{\cal A}_{\mu})
\; +\; \left[ih\; {\mathrm {Tr}}(\overline{H} S \gamma^{\mu}\gamma_5
{\cal A}_{\mu})  \; + \; {\mathrm {h.c.}}\right]
\label{lg}
\end{equation}
by the relation
\begin{equation}
g_{D^{*}D\pi}=\frac{2m_{D}}{f_\pi}g
\end{equation}
valid in the $m_Q \to \infty$ limit.

To compute $g$ in the soft-pion-limit ($q^\mu \to 0$)
we consider the matrix element of $\partial_\mu A^{\mu}$ and derive
a Goldberger-Treiman relation, following the approach of \cite{defazio};
this approach differs from the method employed
in the first paper in ~\cite{ebert}, that assumes
the so-called low-momentum-expansion
approximation and a mixing between the Nambu-Goldstone bosons
and higher mass resonances. Our method amounts to a loop calculation
involving a current and two $H$ states (see Fig. 4). 
\begin{figure}
\epsfxsize=8truecm
\centerline{\epsffile{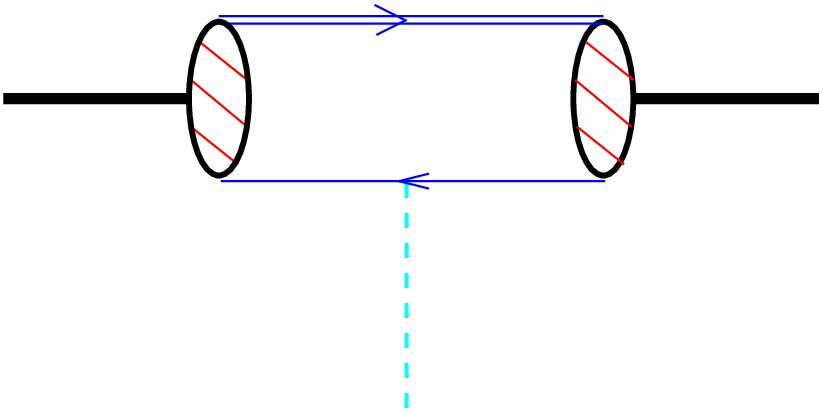}}
\noindent
{\bf Fig. 4} - {Pion vertex on the light quark line for the heavy-heavy-pion
interaction.}
\end{figure}

The result for $g$ is as follows:
\begin{equation}
g=Z_H\left[\frac{1}{3} I_3(\Delta_H)
-2(m+\frac{1}{3}\Delta_H)(I_2 + \Delta_H
I_4(\Delta_H))-\frac{4}{3}m^2 I_4(\Delta_H)\right]~,
\end{equation}
where $I_2,~I_3$ have been defined already and $I_4$ is in the Appendix.
Numerically we get
\begin{equation}
g=0.456\pm0.040
\end{equation}
where the central value corresponds to $\Delta_H=0.4$ GeV and the
lower (resp. higher) value corresponds to $\Delta_H=0.3$ GeV
(resp. $\Delta_H=0.5$ GeV).
These values agree with  QCD sum rules
calculations, that give $g=0.44\pm0.16$ (for a review see
\cite{report}; see also \cite{colg}), 
with the result of relativistic constituent quark model:
$g\simeq 0.40$ \cite{defazio}, $g=0.34$ \cite{gholdom}.
~From the computed values of $g$ we can derive the hadronic width using
\begin{equation}
\Gamma(D^{*+} \to D^0 \pi^+ )  =  \frac{g^2}{6\pi f_\pi^2}
|{\vec p_\pi}|^3~.\label{strong}
\end{equation}
The numerical results will be discussed at the end of this Section.
Let us now comment on the strong decay  $S\to H \pi$. A complete
calculation of the corresponding decay constant $h$  appearing in
(\ref{lg}) is much more involved and one can preliminary
try to compute it in the soft-pion limit. This approximation implies
that we must assume  $\Delta_S = \Delta_H$; putting
 $\Delta=(\Delta_H+\Delta_S)/2$, we obtain
\begin{equation}
h = \overline{Z}\left\{
I_3(\Delta) + 2\Delta I_2 + 2(\Delta^2 -m^2)I_4(\Delta)
\right\}\label{hsp}
\end{equation}
were $\overline{Z}$ is given by
\begin{equation}
\overline{Z}= \left[ \left( I_3(\Delta)+
\Delta\frac{\partial I_3}{\partial \Delta}
\right)^2 - \left( m \frac{\partial I_3}{\partial \Delta}\right)^2
\right]^{-\frac{1}{2}}.
\end{equation}

We obtain, for $\Delta$ in the range $0.43-0.57$ GeV the result
\begin{equation}
h =  -0.85\pm 0.02
\end{equation}
which is somehow higher, but still compatible, within the theoretical
uncertainties, with a result obtained by QCD sum rules:
$h=-0.52 \pm 0.17$ \cite{QCD}.

\subsection{Radiative decays}
Let us now consider the radiative decays $D^* \to D \gamma$
and $B^*\to B \gamma$. To be
definite we consider only the former decay; moreover
we use the SU(3) flavor symmetry for the light quarks.
The matrix element for this radiative transition is the following:
\begin{equation}
{\cal M}(D^* \to D \gamma)
= \hskip 3pt i\,   e \mu \;
\epsilon^{\mu \nu \alpha \beta}\;
\epsilon^*_\mu \; \eta_\nu \; p^{\prime}_\alpha p_\beta
\end{equation}
where $\epsilon_\mu$ is the photon polarization and
the coupling $\mu$ comprises two terms:
\begin{equation}
\mu =  \mu^{\ell}   +  \mu^h ~~ , \label{5.1.5.bis}
\end{equation}
corresponding to the decomposition:
\begin{eqnarray}
{\cal M}(D^* &\to & D \gamma) = e \; \epsilon^{* \mu} \;
<D(p^{\prime})|J_\mu^{em}|D^* (p, \eta)> \nonumber \\
& = & e \; \epsilon^{* \mu} \;
<D(p^{\prime})|J_\mu^{\ell}+ J_\mu^{h}|D^* (p, \eta)>~~.
\label{5.1.3}
\end{eqnarray}
Here $J_\mu^{\ell}$ and $J_\mu^{h}$ are the light and
the heavy quark parts of the electro-magnetic current:
\begin{equation}
J_\mu^{\ell}=
\frac{2}{3} \bar{u} \gamma_\mu u -
\frac{1}{3} \bar{d} \gamma_\mu d -
\frac{1}{3} \bar{s} \gamma_\mu s
\label{5.1.4}
\end{equation}
and
\begin{equation}
J_\mu^{h}=
\frac{2}{3} \bar{c} \gamma_\mu c -
\frac{1}{3} \bar{b} \gamma_\mu b =  \sum_{Q=c,b} e_Q
\bar{Q} \gamma_\mu Q.
\label{5.1.5}
\end{equation}
Correspondingly, eq. (\ref{5.1.5.bis}) becomes
\begin{equation}
\mu = \mu^{\ell} + \mu^h =
\frac{e_q}{\Lambda_q} + \frac{e_Q}{\Lambda_Q},
\label{5.1.5.ter}
\end{equation}
where $\Lambda_q$ and $\Lambda_Q$ are mass parameters to be determined.
The photon insertion on the heavy quark line
generates the $\mu^h$ coupling, while the $\mu^{\ell}$ arises when the photon
is inserted on the light quark line of the loop.

In the $m_Q\to\infty$ limit,
the matrix element of $J_\mu^h$  can be expressed in terms of the
Isgur-Wise  form factor $\xi(\omega)$ as follows:
\begin{eqnarray}
<D(p^{\prime})| &J_\mu^h& |D^*(p, \eta)> = \, e_c
<D(p^{\prime})|\bar{c} \gamma^\mu c|D^*(p, \eta)> \nonumber \\
& = & i \frac{2}{3}  \sqrt{M_{D} M_{D^*}} \xi(\omega)
\epsilon_{\mu \nu \alpha \beta}
\eta^\nu v^{\prime \alpha} v^{\beta},
\label{5.1.6}
\end{eqnarray}
where $p^{\prime}=M_D v^{\prime}$, $p=M_{D^*} v$ and $\omega\simeq 1$ because:
\begin{equation}
0=q^2=m^2_D + M_{D^*}^2 -2 M_D M_{D^*} v \cdot v^{\prime}.
\label{5.1.7}
\end{equation}

Taking into account the normalization $\xi(1)=1 $ one gets,
for the charm and beauty mesons respectively (with $\bar q = \bar d$),
\begin{equation}
\mu^h \; = \; \frac{2}{3 \Lambda_c}~~~~~~~
\mu^h \; = \; - \; \frac{1}{3 \Lambda_b}
\label{5.1.8}
\end{equation}
with
\begin{equation}
\Lambda_c={\sqrt{ M_{D} M_{D^*}}}~~~~~~~~~~~
\Lambda_b={\sqrt{ M_{B} M_{B^*}}}~~.
\label{lambdac}
\end{equation}
In the leading order in $1/m_c$ and $1/m_b$ one finds:
\begin{equation}
\Lambda_c\; =\; m_c~~~~~~~
\Lambda_b\; =\; m_b
\label{5.1.8sec}
\end{equation}
which fixes $\mu^h$. As for the coupling $\mu^{\ell}$,
in our approach it stems from the diagram
where the photon line is inserted on the light quark propagator (see Fig. 5).
\begin{figure}
\epsfxsize=8truecm
\centerline{\epsffile{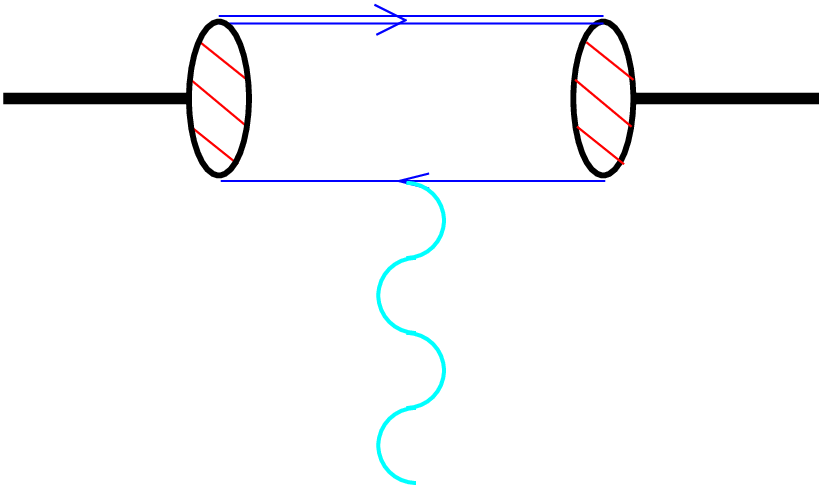}}
\noindent
{\bf Fig. 5} - {Electro-magnetic current insertion on the light quark line.}
\end{figure}

The result is:
\begin{equation}
\beta  = \Lambda_q^{-1}=2 \times Z_H [I_2 +  (m + \Delta_H) I_4(\Delta_H)]
\end{equation}
Numerically we get:
\begin{equation}
\beta=1.6\pm 0.1~{\mathrm {GeV}}^{-1}
\end{equation}
where the central value corresponds to $\Delta_H=0.4$ GeV and the
lower (resp. higher) value corresponds to $\Delta_H=0.5$ GeV
(resp. $\Delta_H=0.3$ GeV).
These values agree with the result of Heavy Quark Effective Theory and the 
Vector Meson Dominance hypothesis \cite{defazio2}.

One can use the formula:
\begin{equation}
\Gamma(D^* \to D \gamma ) =  \frac{\alpha}{3}
\frac{M_{D^*}}{M_{D}}
|\mu |^2 |{\vec k}|^3 \label{gg}
\end{equation}
($ {\vec k}$ = photon momentum), to compute the $D^*$ radiative widths.
Using this equation together with (\ref{strong})
we obtain the results of Table VII.
\begin{table} [htb]
\hfil
\vbox{\offinterlineskip
\halign{&#& \strut\quad#\hfil\quad\cr
\tableline
\tableline
&Decay mode&& $\Delta_H=0.4$ GeV && $\Delta_H=0.5$ GeV && Exp.&\cr
\tableline
&${D^*}^0\to D^0 \pi^0$&& $65.5$ && $70.1$ && $61.9\pm 2.9$&\cr
&${D^*}^0\to D^0 \gamma$&& $34.5$ && $29.9$ &&$38.1\pm 2.9$&\cr
&${D^*}^+\to D^0 \pi^+$&& $71.6$ && $71.7$ && $68.3\pm 1.4$&\cr
&${D^*}^+\to D^+ \pi^0$&& $28.0$ && $28.1$ && $30.6\pm 2.5$&\cr
&${D^*}^+\to D^+ \gamma$&& $0.4$ &&  $0.24$ &&$1.1^{+2.1}_{-0.7}$&\cr
\tableline
\tableline}}
\caption{Theoretical and  experimental $D^*$ branching ratios (\%).
Theoretical values are computed with $\Delta_H=0.4,~0.5$ GeV.}
\end{table}
Taking into account the approximations involved in the present calculation,
we find comparison between theoretical prediction and experimental data 
of Table VII encouraging. Finally we compute total widths for $D^*$ and $B^*$
(for $\Delta_H = 0.4$ GeV):
\begin{eqnarray}
\Gamma (D^{*0})&=& 38 \;{\mathrm {KeV}} \\
\Gamma (D^{*+})&=& 62 \;{\mathrm {KeV}}  \\
\Gamma (B^{*0} \to B^0\gamma) &=& 0.05 \;{\mathrm {KeV}}
\end{eqnarray}

\section{Conclusions}

The increasing number of available data on heavy meson processes and the even
more promising increase of data from forthcoming experiments demands
theoretical predictions for these processes to be compared with experiment and
to suggest subsequent lines of investigation. Calculating directly from the QCD
lagrangian remains an extremely difficult task, in spite of the impressive
success of recent lattice work. A most promising approach is the one based
on heavy meson effective lagrangians, which incorporate the heavy quark
symmetries
and in addition the approximate chiral symmetry for light quarks. Although with
increasing data such an approach will remain the best one beyond direct QCD
calculations, at this stage it is made cumbersome by the large number of
parameters that have to be fixed before obtaining predictions.
In this note we have presented an intermediate approach, not as rigorous
and general as that of the effective meson lagrangian,
but which allows for a smaller number of input parameters.
We start from an effective lagrangian at the level of mesons and of
constituent
quarks, and we then calculate the
meson transition amplitudes by evaluating loops of heavy and light quarks.
In this way we can compute the Isgur-Wise function, the form factors
$\tau_{1/2}$ and $\tau_{3/2}$, the leptonic decay constant ${\hat F}$ and
${\hat F}^+$, the coupling constants $g$ and $h$ relative to $H \to H\pi$
and $H\to S\pi$ processes respectively, the $\beta$ coupling relative to the
$H\to H\gamma$ processes. The agreement with data, when available, seems 
rather impressive. Additional data should be
able to confirm some of the predictions of the model or
suggest modifications. In conclusion the model presented here,
based on meson and quark degrees of freedom, seems capable of incorporating
a number of essential features of QCD and
to provide a useful approach to calculate heavy meson transitions in terms
of a limited number of physical parameters.

\newpage
\section{Appendix}
We list here the integrals used in the paper.
\begin{eqnarray}
I_0(\Delta)&=& \frac{iN_c}{16\pi^4} \int^{\mathrm {reg}}
\frac{d^4k}{(v\cdot k + \Delta + i\epsilon)} \nonumber \\
&=&{N_c \over {16\,{{\pi }^{{3/2}}}}}
\int_{1/{{{\Lambda}^2}}}^{1/{{{\mu }^2}}} {ds \over {s^{3/2}}}
\; e^{- s( {m^2} - {{\Delta }^2} ) }
\left( {3\over {2\,s}} + {m^2} - {{{\Delta}}^2} \right)
[1+{\mathrm {erf}}(\Delta\sqrt{s})]\nonumber \\
&-& \Delta {{N_c m^2}\over {16 \pi^2}} \Gamma(-1,{{{m^2}}
\over {{{\Lambda}^2}}},{{{m^2}}\over {{{\mu }^2}}})
\\
I_1&=&\frac{iN_c}{16\pi^4} \int^{reg} \frac{d^4k}{(k^2 - m^2)}
={{N_c m^2}\over {16 \pi^2}} \Gamma(-1,{{{m^2}}
\over {{{\Lambda}^2}}},{{{m^2}}\over {{{\mu }^2}}})
\\
I_1^{\prime}&=&\frac{iN_c}{16\pi^4} \int^{\mathrm {reg}} d^4
k\frac{k^2}{(k^2 - m^2)}
={{N_c m^4}\over {8 \pi^2}} \Gamma(-2,{{{m^2}}
\over {{{\Lambda}^2}}},{{{m^2}}\over {{{\mu }^2}}})\\
I_2&=&-\frac{iN_c}{16\pi^4} \int^{\mathrm {reg}}  \frac{d^4k}{(k^2 - m^2)^2}=
\frac{N_c}{16\pi^2} \Gamma(0,\frac{m^2}{\Lambda^2}, \frac{m^2}{\mu^2})\\
I_3(\Delta) &=& - \frac{iN_c}{16\pi^4} \int^{\mathrm {reg}}
\frac{d^4k}{(k^2-m^2)(v\cdot k + \Delta + i\epsilon)}\nonumber \\
&=&{N_c \over {16\,{{\pi }^{{3/2}}}}}
\int_{1/{{\Lambda}^2}}^{1/{{\mu }^2}} {ds \over {s^{3/2}}}
\; e^{- s( {m^2} - {{\Delta }^2} ) }\;
\left( 1 + {\mathrm {erf}} (\Delta\sqrt{s}) \right)\\
I_4(\Delta)&=&\frac{iN_c}{16\pi^4}\int^{\mathrm {reg}}
\frac{d^4k}{(k^2-m^2)^2 (v\cdot k + \Delta + i\epsilon)} \nonumber\\
&=&\frac{N_c}{16\pi^{3/2}} \int_{1/\Lambda^2}^{1/\mu^2} \frac{ds}{s^{1/2}}
\; e^{-s(m^2-\Delta^2)} \; [1+{\mathrm {erf}}(\Delta\sqrt{s})]~.
\end{eqnarray}
In these equations
\begin{equation}
\Gamma(\alpha,x_0,x_1) = \int_{x_0}^{x_1} dt\;  e^{-t}\; t^{\alpha-1}
\end{equation}
is the generalized incomplete gamma function and erf is the error function.

Let's introduce now:
\begin{equation}
\sigma(x,\Delta_1,\Delta_2,\omega)={{{\Delta_1}\,\left( 1 - x \right)  +
{\Delta_2}\,x}\over {{\sqrt{1 + 2\,\left(\omega -1  \right) \,x +
2\,\left(1-\omega\right) \,{x^2}}}}}~.
\end{equation}
Then the integrals used for the semileptonic decays are as follows:
\begin{eqnarray}
I_5(\Delta_1,\Delta_2,\omega) & = & \frac{iN_c}{16\pi^4} \int^{\mathrm {reg}}
\frac{d^4k}{(k^2-m^2)(v\cdot k + \Delta_1 + i\epsilon )
(v'\cdot k + \Delta_2 + i\epsilon )} \nonumber \\
 & = & \int_{0}^{1} dx \frac{1}{1+2x^2 (1-\omega)+2x
(\omega-1)}\times\nonumber\\
&&\Big[ \frac{6}{16\pi^{3/2}}\int_{1/\Lambda^2}^{1/\mu^2} ds~\sigma
\; e^{-s(m^2-\sigma^2)} \; s^{-1/2}\; (1+ {\mathrm {erf}}
(\sigma\sqrt{s})) +\nonumber\\
&&\frac{6}{16\pi^2}\int_{1/\Lambda^2}^{1/\mu^2}
ds \; e^{-s(m^2-2\sigma^2)}\; s^{-1}\Big]
\end{eqnarray}

\begin{eqnarray}
I_6(\Delta_1,\Delta_2,\omega)&=&\frac{iN_c}{16\pi^4} \int^{\mathrm {reg}}
\frac{d^4k}{(v\cdot k + \Delta_1 + i\epsilon)
(v'\cdot k + \Delta_2 + i\epsilon)} \nonumber \\
&=&I_1 \int_{0}^{1} dx
\frac{\sigma}{1+2x^2(1-\omega) + 2x(\omega-1)}\nonumber\\
&-&\frac{N_c}{16 \pi^{3/2}}\int_{0}^{1} dx
\frac{1}{1+2x^2(1-\omega) + 2x(\omega-1)}\times\nonumber\\
&&\int_{1/\Lambda^2}^{1/\mu^2} \frac{ds}{s^{3/2}}
e^{-s(m^2-\sigma^2)}\Big\{\sigma
[1+{\mathrm {erf}}(\sigma\sqrt{s})]\cdot[1+2s(m^2-\sigma^2)]\nonumber\\
& +&2 {\sqrt{\frac{s}{\pi}}} e^{-s\sigma^2}\left[
\frac{3}{2s} + (m^2-\sigma^2)\right]
\end{eqnarray}

\begin{eqnarray}
S(\Delta_1,\Delta_2,\omega)&=&\Delta_1\,I_3(\Delta_2) + \omega \,
\left( I_1 + \Delta_2 \,I_3(\Delta_2 ) \right)  +
{{\Delta_1}^2}\,I_5(\Delta_1,\Delta_2 ,\omega ) \nonumber \\
T(\Delta_1,\Delta_2,\omega)&=&{m^2}\,I_5(\Delta_1,\Delta_2 ,\omega ) +
I_6(\Delta_1,\Delta_2 ,\omega ) \nonumber \\
U(\Delta_1,\Delta_2,\omega)&=&I_1 + \Delta_2 \,I_3(\Delta_2 ) + \Delta_1\,
I_3(\Delta_1) + \Delta_2 \,\Delta_1\,I_5(\Delta_1,\Delta_2 ,\omega )
\end{eqnarray}

\twocolumn
\par\noindent
\vspace*{1cm}
\par\noindent
{\bf Acknowledgments}
\par\noindent
A.D. acknowledges the support of a ``Marie Curie'' TMR research fellowship
of the European Commission under contract ERBFMBICT960965. He also thanks
L. Lellouch for a discussion on lattice data and O. P\`ene for discussion
on ref. \cite{oliver}. This work has also been carried out in part under the
EC program Human Capital and Mobility, contract UE
ERBCHRXCT940579 and OFES 950200.

\end{document}